\documentstyle[11pt,paspws98,psfig]{article}
\begin{document}
\title{Radiative Transfer Effects and Convective Collapse: Size(flux)-Strength 
Distribution for the Small-scale Solar Magnetic Flux Tubes} 
\author{S.Rajaguru\footnote{Joint Astronomy Programme, Indian Institute of Science, Bangalore} 
                \& S.S.Hasan}
\affil{Indian Institute of Astrophysics, Bangalore-560 012, India }
\begin{abstract}
The effects of radiative energy exchange on the convective instability
of a weak field magnetic structure, which lead to a prediction and a
physical explanation of the magnetic flux dependent field strength, are
examined in detail using a real model stratification for the photospheric
and convection zone structure of the Sun. Adopting the generalised
Eddington approximation for the radiative transfer, which is valid both 
in the optically thick and the thin limits, we model the lateral radiative
energy exchange by the tube with the external medium with a self-consistent
inclusion of vertical radiative losses.
\end{abstract}
 
\section{Introduction}
One of the important properties of the small-scale solar magnetic structures,
as established by the observations(Stenflo and Harvey 1985,
Zayer et al 1989,1990; Schussler 1991), 
is that while the strong field $network$ $elements$
have field strengths very weakly dependent on the flux per element, $\Phi$, with typical
values of $\Phi$ about $1\times 10^{18}$ Mx and higher, the $inner$ $network$
weak field structures have a typical value of about $500$ G(Keller et al 1994, Solanki et al 1996,
Lin 1995) for their
strength and $\Phi$ about $1\times 10^{17}$ Mx or lower with a strong dependence between
the two.  
If the convective
collapse of a weak field tube is the global process responsible for the formation of
the strong field tubes(Parker 1978; Webb and Roberts 1978; Spruit and Zweibel 1979;
Spruit 1979; Hasan 1983, 1984;Venkatakrishnan 1985; Steiner 1996), 
that comprise the $network$, with strengths weakly dependent on 
$\Phi$ then it should be explained why tubes with smaller fluxes viz., the $inner$ 
$network$ elements do not collapse to kG strength; efficient radiative exchange with the
surroundings by a small flux tube(Hasan 1986; Venkatakrishnan 1986) offers a natural
explanation; 
here, having in mind a quantitative comparison with the above mentioned observationally
established properties of the Solar flux tubes, we employ a semi-empirical model
of the photospheric and the convection zone structure of the Sun and study in detail the
effects of radiation on the convective instability and the wave motions.
 
\section{Equations}
We add to the familiar thin flux tube equations(Roberts and Webb,1978) the following
non-adiabatic energy equation(see e.g.,Cox 1980),
\begin{equation}
\frac{\partial p}{\partial t}+v\frac{\partial p}{\partial z}-\frac{\Gamma_{1} p}{\rho}\left [
\frac{\partial \rho}{\partial t}+v\frac{\partial \rho}{\partial z}\right ]=
-\frac{\chi_{T}}{\rho c_{v}T}\nabla .\bf F
\label{g.energy}
\end{equation}
where $p$, $\rho$, and ${\bf F}$ denote the fluid pressure, density, and 
radiative energy flux respectively; (see Cox(1980) for the definitions of other thermodynamic
quantities); 
all the variables are evaluated on the tube axis(r=0); 
The radiative flux ${\bf F}$ is calculated in the generalised Eddington
approximation following Unno and Spiegel(1967): the mean intensity $J$ that is needed
to evaluate the flux,
\begin{equation}
{\bf F}_{R}=-\frac{4}{3\kappa \rho}{\bf \nabla}J
\label{Rflux.eddn}
\end{equation}
is found by reducing the exact relation
\begin{equation}
\nabla .{\bf F}_{R}=4\kappa \rho (S-J)
\label{divflux}
\end{equation}
to a form appropriate for a thin tube, which reads,
\begin{equation}
\frac{1}{3}\frac{\partial^{2}J}{\partial \tau^{2}}+\frac{4}{3}\left (\frac{J_{e}-J}{\tau_{a}^{2}}
\right )=J-S
\label{transfer.tube}
\end{equation}
where $d\tau =\kappa \rho dz$, $\tau_{a}=\kappa\rho a$, $S$ is the source function
which we take it to be the Planck function, 
$a$ is the tube radius, and $J_{e}$ is the mean intensity
in the external medium. $J_{e}$ is found by solving the equation
\begin{equation}
\frac{1}{3}\frac{\partial^{2}J_{e}}{\partial \tau^{2}_{e}}=J_{e}-S_{e}
\label{transfer.ext}
\end{equation}

\subsection{Equilibrium}
The equilibrium stratification of the external medium that we use here is the one
determined to match the combined $VAL-C$(Vernazza et al. 1981) and Spruit(1977) models
(see Hasan,Kneer and Kalkofen(1998) for details about the construction of this
external quite Sun model); 
the Rosseland mean opacities are calculated by interpolation
from the tables of Kurucz(1993) for the upper layers and from those of Rogers and Iglesias(1992)
for the deeper regions. 
We assume temperature equilibrium, $T=T_{e}$, which implies that $\beta $,
defined as $\beta = 8 \pi p/B^{2}$, is constant with z(if the dependence of $\mu$ is 
neglected). The pressure and density are thus determined from
\begin{eqnarray} 
p=\frac{\beta}{1+\beta}p_{e} \\
\rho=\frac{\beta}{1+\beta}\rho_{e} 
\end{eqnarray}
The extent of the flux tube 
covers the atmosphere from the temperature minimum in the 
chromosphere to $5000$km deep in the convection zone with the photospheric surface($\tau=1$)
being assigned $z=0$. We measure the positive $z$ downwards. 
\subsection{Linear Stability: The Perturbation Equations}
With the assumption that the perturbations in the ambient medium are negligible
small perturbations inside the tube about the equilibrium described above obey the
following equations in the linear limit:
\begin{equation}
\frac{\partial\xi}{\partial z}=\frac{B^{'}}{B}-\frac{\rho^{'}}{\rho}-\left [\frac{d (\ln\rho)}
{dz}-\frac{d (\ln B)}{dz}\right ]\xi
\end{equation}
\begin{equation}
\frac{\partial^{2}\xi}{\partial t^{2}}=-Hg\frac{\partial}{\partial z}\left (\frac{p^{'}}
{p}\right )-g\left (\frac{p^{'}}{p}-\frac{\rho^{'}}{\rho}\right )
\end{equation}
\begin{equation}
\frac{B^{'}}{B}=-\frac{\beta}{2}\frac{p^{'}}{p}
\end{equation}
\begin{equation}
\frac{\partial}{\partial t}\left (\frac{p^{'}}{p}\right )-\Gamma_{1}\frac{\partial}{\partial t}
\left (\frac{\rho^{'}}{\rho}\right )+\left [\frac{d\ln p}{dz}-\Gamma_{1}\frac{d\ln\rho}{dz}
\right ]\frac{\partial \xi}{\partial t}=-\frac{\chi_{T}}{\rho c_{v}T}\nabla .{\bf F^{'}}
\label{p.energy}
\end{equation}
where $\xi$ denotes the vertical displacement and $H$ is the pressure scale height.
the perturbation of equation (\ref{divflux})
can be done in a straightforward manner:
\begin{equation}
\nabla.{\bf F_{R}^{'}}=\frac{dF_{R}^{'}}{dz}=4\pi\kappa_{a}\rho(S^{'}-J^{'})+
\nabla.{\bf F_{R}}\left (\frac{\kappa^{'}}{\kappa}+\frac{\rho^{'}}{\rho}\right )
\label{divflux.p}
\end{equation}
The perturbation in the mean intensity $J^{'}$ is determined by perturbing and linearizing
the transfer equation (\ref{transfer.tube}) which in the first order moment form are 
\begin{equation}
\frac{d}{dz}\left (\frac{J^{'}}{J}\right )=\frac{dlnJ}{dz}\frac{\eta^{'}}{\eta}-
\frac{dlnJ}{dz}\frac{J^{'}}{J}+\frac{dlnJ}{dz}\frac{{\mathcal{H}}^{'}}{\mathcal{H}}
\label{p.meani}
\end{equation}
\begin{eqnarray}
\frac{d}{dz}\left (\frac{{\mathcal{H}}^{'}}{\mathcal{H}}\right )=\left (\frac{\Delta_{t}}{2
\epsilon H_{p}}-\frac{dln{\mathcal{H}}}{dz}\right )\frac{\eta^{'}}{\eta}+
\frac{\Delta_{t}}{2\epsilon H_{p}}\frac{a^{'}}{a}-
\frac{(1+\nabla_{c})(4+3\tau_{a}^{2})}{16\epsilon H_{p}}\frac{J^{'}}{J}+ \nonumber \\
\frac{1}{qH_{p}}\frac{T^{'}}{T}-\frac{dln{\mathcal{H}}}{dz}\frac{{\mathcal{H}^{'}}}{\mathcal{H}}
\label{p.eddnflx}
\end{eqnarray}

where ${\mathcal{H}}={\bf F}/4$ is the Eddington flux, $\eta=\kappa \rho$, $H_{p}$ is the pressure 
scale-height at the bottom and the various other quantities are as defined below:
\begin{equation}
\nabla_{c}=\frac{J}{S}-1
\end{equation}
is a measure of departure from radiative equilibrium, 
\begin{equation}
\Delta_{t}=\frac{J-J_{e}}{S}
\end{equation}
is the ratio of the excess of the mean intensity inside the tube to the Planck function;
$\epsilon$ and $q$ are the ratios,
\begin{equation}
\epsilon=\frac{\tau_{r}}{\tau_{th}}
\end{equation}
\begin{equation}
q=\frac{\tau_{N}}{\tau_{th}}
\end{equation}
where 
\begin{equation}
\tau_{th}=\frac{\rho c_{v}TH_{p}}{\mathcal{H}}
\end{equation}
is the radiative relaxation time over the length of one pressure scale-height at the
bottom,
\begin{equation}
\tau_{r}=\frac{\rho c_{v}a^{2}}{K},     K=\frac{16\sigma T^{3}}{3\kappa \rho}
\end{equation}
is the radiative relaxation time across the tube in the optically thick limit and
\begin{equation}
\tau_{N}=\frac{c_{v}T}{4\kappa S}
\end{equation}
is the radiative relaxation time that one obtains in the optically thin limit
(with $Newton's$ $law$ $of$ $cooling$)(Spiegel,1957); and,
\begin{equation}
\tau_{a}=\kappa \rho a
\end{equation}
is the depth dependent optical thickness of the tube.
We reduce the perturbation equations to a
final set of four equations for the four variables $\xi$, $p^{'}/p$, $J^{'}/J$ and
${\mathcal{H}}^{'}/{\mathcal{H}}$. 
%It is easy to check that the equations
%get reduced to the correct forms for the optically thick and optically thin limits. The
The optically thick reduction of the set of equations corresponds to taking the limit $\tau$ 
tending to infinity and replacing the mean intensity by the Planck function. The optically
thin case is got in the limit $\tau_{th}$ approaching zero. 
%This proven validity of
%the generalised Eddington approximation(Unno and Spiegel 1967) for both the limits makes
%the analysis of the present problem realistic.
\subsection{Boundary Conditions}
We use closed mechanical boundary conditions and as thermal conditions we impose that there
is no incoming radiation from above at the top boundary and that the perturbations are
adiabatic at the bottom boundary. 
\section{Results and Discussion}
The action of radiation that we bring out in this study of convective instability and 
wave-motions of the tube is explained conveniently with the help of the graphs shown in
$figs.(1)$ and $(2)$ where we plot the growth rates and frequencies of the fundamental mode,
which is the most unstable, respectively, as a function of the surface tube radius $a_{0}$
for various values of the plasma $\beta$.
 
\begin{figure}
\centerline{\psfig{file=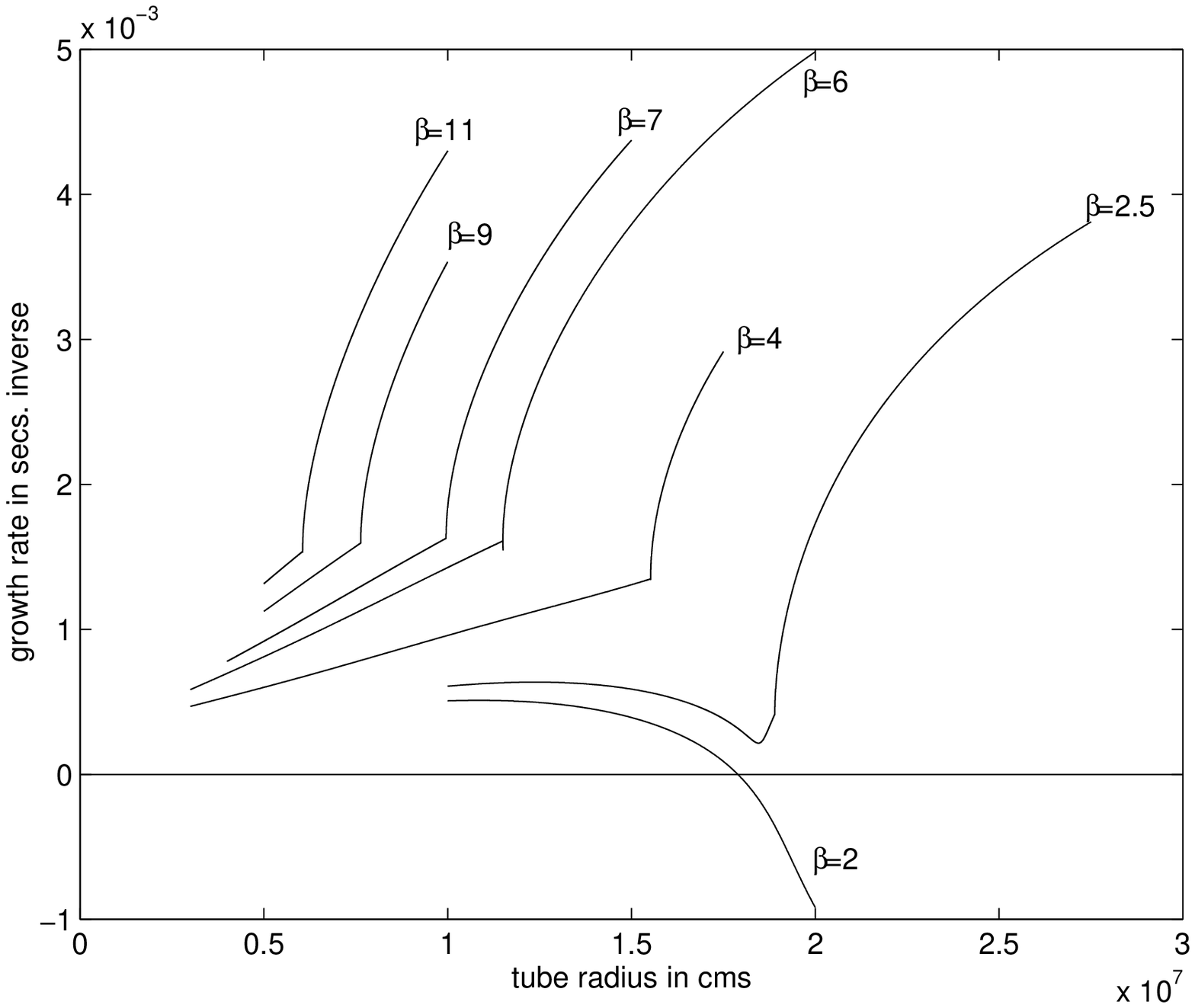,width=8.5cm}}
\label{fig.(4.5)}
\caption{Growth rates of the fundamental mode as a function of the tube radius $a_{0}$
at the surface $z=0$($\tau=1$) for various values of the plasma $\beta$.}
\vskip 0.5in
\centerline{\psfig{file=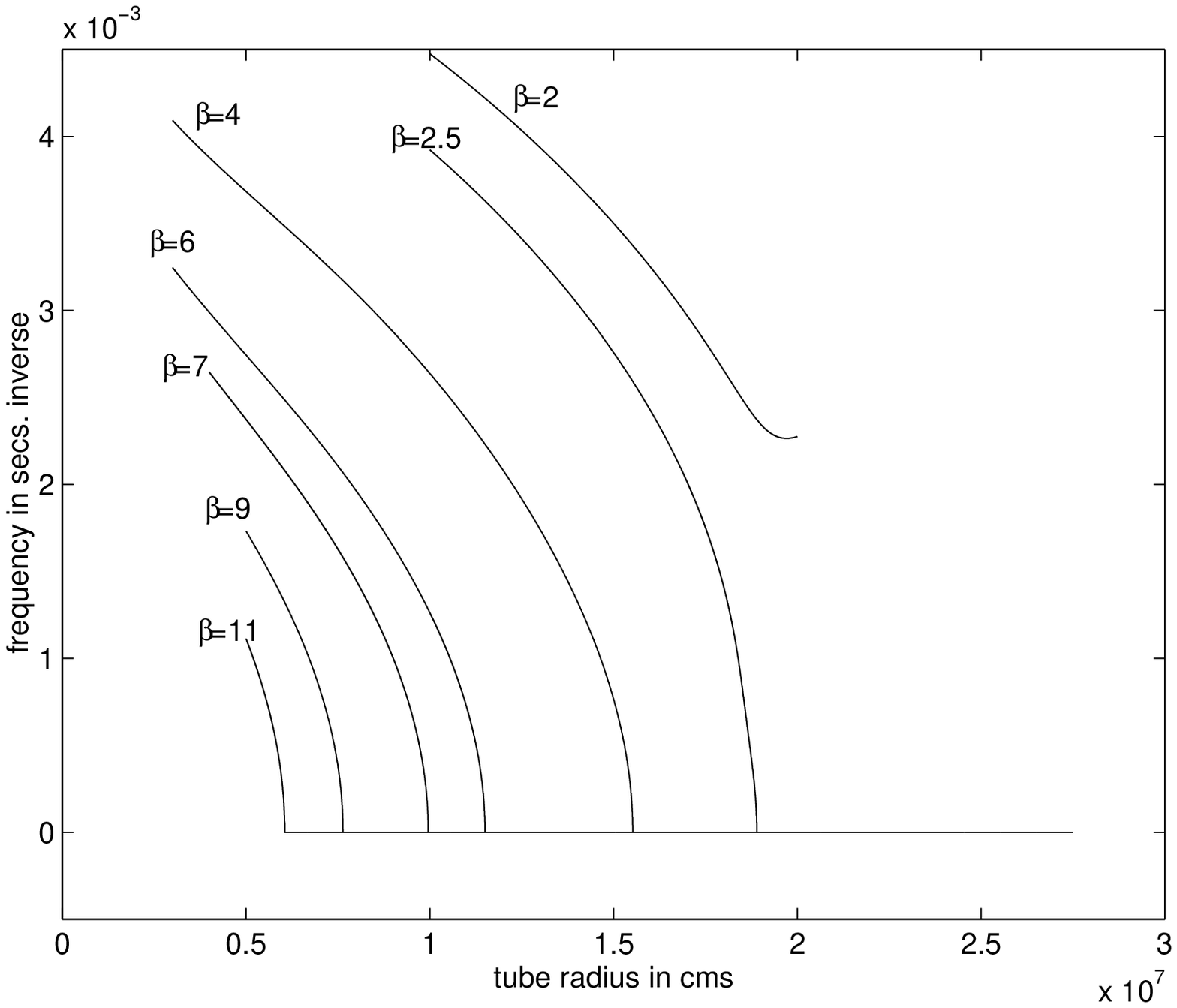,width=8.5cm}}
\label{fig.(4.6)}
\caption{The frequencies of the fundamental mode as a function of the tube radius $a_{0}$
at the surface $z=0$($\tau=1$) for various values of the plasma $\beta$.}
\end{figure}
\subsection{Convective Instability} 
The onset of the convective instability corresponds to the cusps in the curves of $fig.(1)$ 
where the overstable mode's frequency becomes zero and the growth rate shoots up sharply. 
Comparison with the
results obtained with only the lateral exchange in the diffusion approximation
(Venkatakrishnan
1986, Hasan 1986) or in the Newton's law of cooling reveal that these earlier treatments 
overestimate the degree of instability: the growth rates obtained with the present more accurate
treatment of radiation with the inclusion of vertical exchange of 
radiation are appreciably smaller; 
moreover, for a given value of $\beta $, i.e.
for a tube of given strength the onset of convective instability requires the size of the tube
to be greater in the present case than that is required when the diffusion approximation 
is used.
 
 The convective instability is completely suppressed for tubes with the plasma $\beta$ smaller 
than $2.45$ whatever be its size; this corresponds to a field strength of about $1160$Gauss at
$\tau=1$ inside the tube; 
this has to be compared with the value of $1350$G($\beta=1.83$) that Spruit and Zweibel
obtained in the adiabatic case. 
We point out here that the field 
strength of $1160$G that we obtain here does not necessarily imply that all collapsing tubes of
weaker fields will attain this unique value and become stable; this value represents a necessary
strength for stability against convective collapse and thus can be considered as a minimum
strength for stability; a collapsing weaker tube of sufficient size can of course attain an
equilibrium collapsed state of field strength higher than this value(cf. Spruit 1979). 
 
\begin{figure}
\centerline{\psfig{file=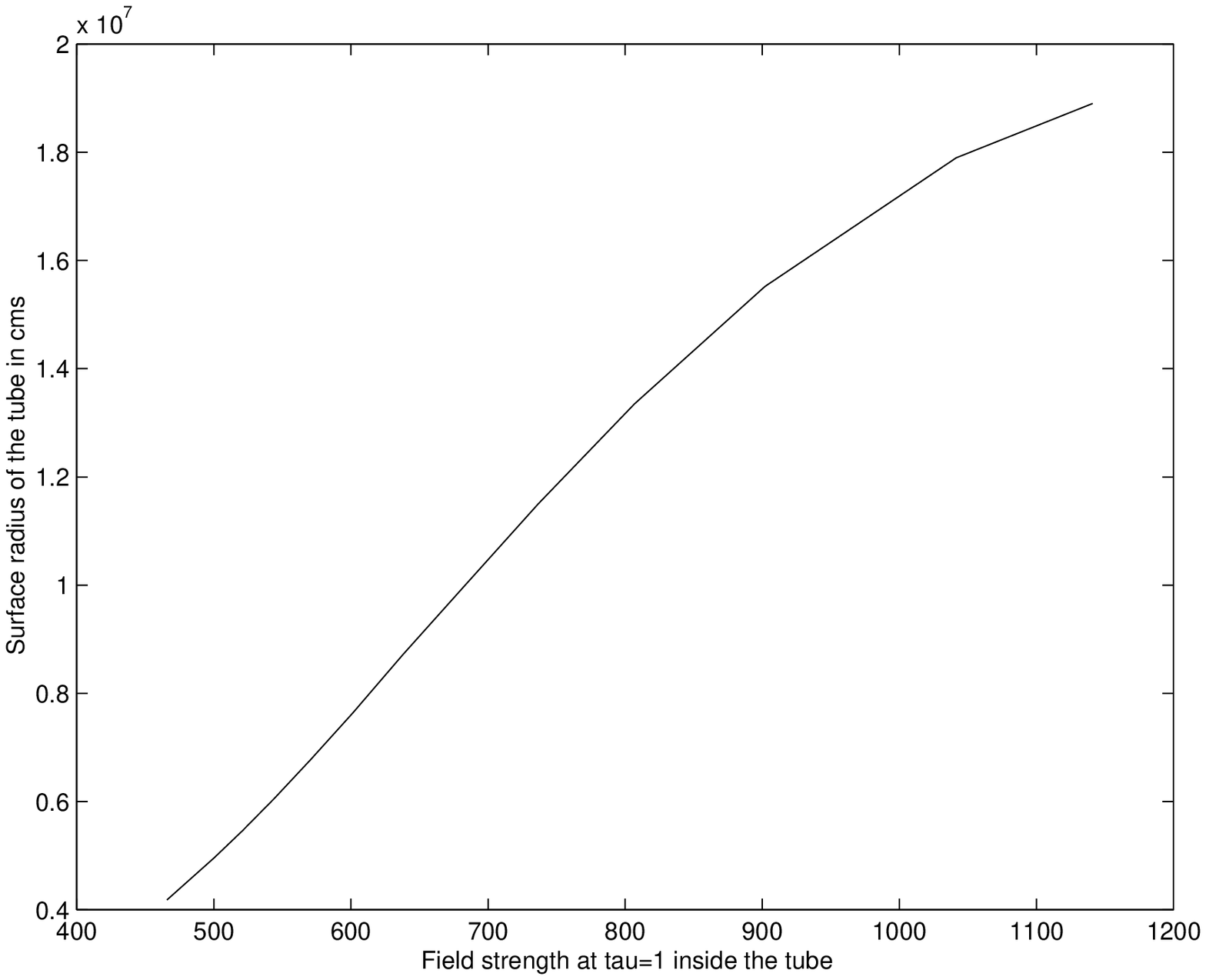,width=8.5cm}}
\caption{Size-strength relation for the stable tubes; see the text for details}
\vskip 0.5in
\centerline{\psfig{file=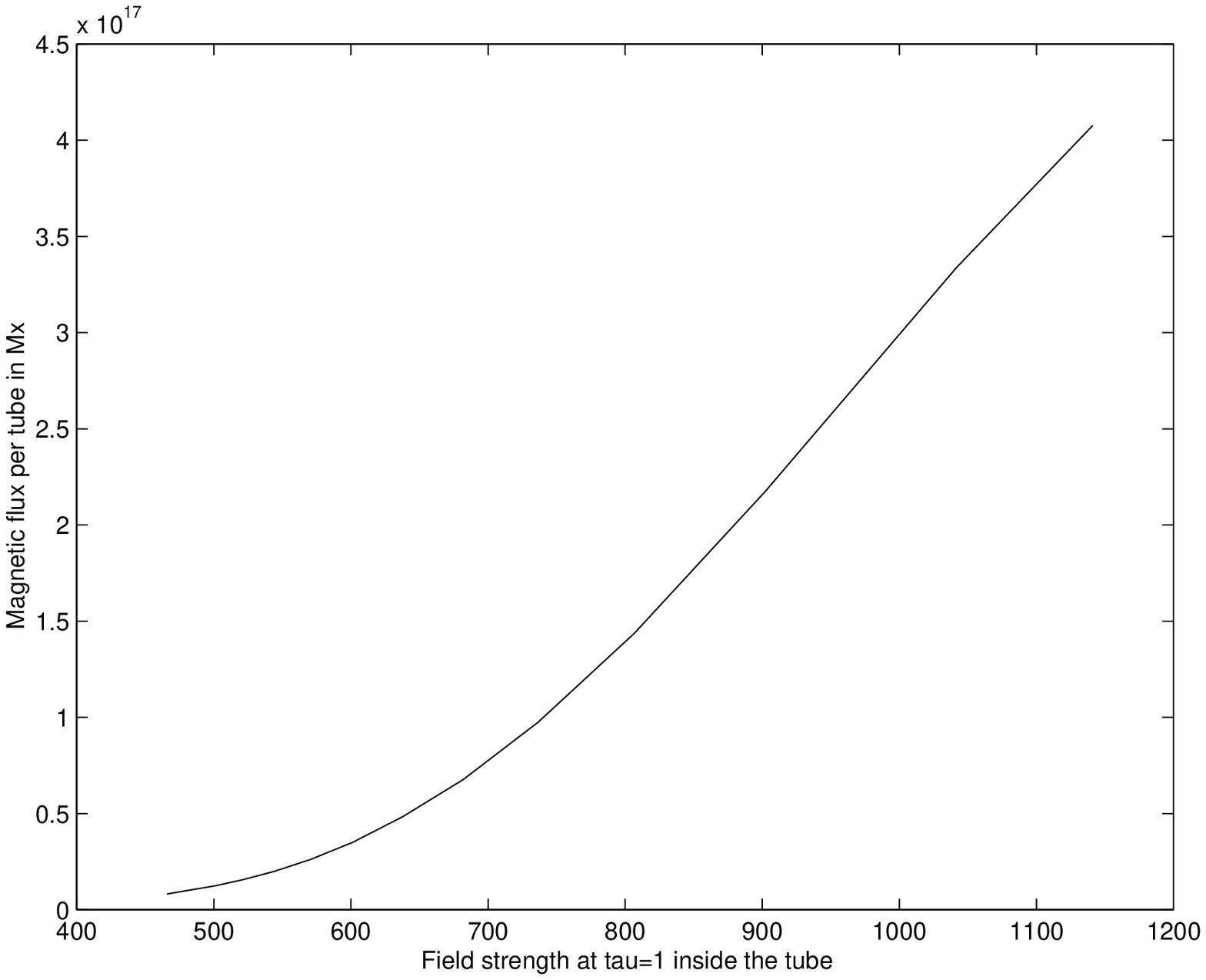,width=8.5cm}}
\caption{The magnetic flux-strength relation corresponding to the size-strength relation of
the previous graph}
\end{figure} 
\subsection{Size-strength Distribution for the Solar Tubes}
From the positions of the cusps in the curves of $fig.(1)$, i.e. from the positions
that mark the onset of the convective instability we pick up the values of the plasma $\beta$
and the radius $a_{0}$; the resulting radius-field strength dependence is shown in $fig.(3)$;
the corresponding flux-strength distribution
in $fig.(4)$. Comparison of this curve with those observationally produced(Solanki et al 1997,
Lin 1996) shows a remarkable agreement leading to the conclusion that indeed the convective
collapse is the cause of the formation of the flux elements on the Sun's surface. Our refined,
realistic, and exact treatment of the convective collapse process on the Sun reinforces the
conclusions drawn from earlier simplified treatments(Venkatakrishnan 1986, Hasan 1986) and 
verifies the original suggestion and the physical explanation by Parker(1978) for the 
concentrated small scale magnetic structures on the Sun.
\subsection{Overstability} 
 The characteristics of the overstable mode are explained with the help of the $fig.(1)$ again;
the growth rates in the present Eddington approximation are lower than those
obtained in the earlier treatments where the vertical losses are not taken care of
and use either the diffusion approximation or the Newton's law of cooling; 
the differences thus demonstrate that overstability is hindered by the vertical losses

 We point out that while there is no damping out of oscillations when only lateral
exchange takes place and the oscillations' growth rate only asymptotically goes to zero
in the limit of large radii, i.e. in the adiabatic limit, the inclusion of vertical
radiative losses make the oscillations to damp out for radii greater than a particular finite
value which is determined by the $\beta$; thus it is clarified
that the horizontal exchange between vertically oscillating fluid elements acts to amplify
the oscillations while the vertical losses always try to smooth out the fluctuations thereby
introducing damping; this $radiative$ $damping$ is quite severe for the overstable mode of an
intense flux tube on the Sun that it gets completely damped out for tubes of radii larger than
a certain critical value.
 
 Finally we note that a tube which can undergo convective
collapse for radii greater than a critical value for a given field strength remains overstable
for all smaller radii that it can take.  
\section{Conclusions}
\begin{itemize}
\item
We have demonstrated that radiative transport has a marked effect on the
size-field strength relation for solar flux tubes. Our results can be
be applied to solar flux tubes more reliably in view of the more
refined treatment of radiative transfer.
\item
We have generalized the necessary condition for the onset of convective
instability in the presence of radiative energy exchange. We find that radiation
has a stabilizing influence which is greater for tubes with small radius.
\item
Overstability of the longitudinal slow mode is shown to be dependent on the
tube radius: there is a critical tube radius above which a strong field convectively
stable tube's oscillations get damped as a result of vertical radiative losses. 
\end{itemize}

\end{document}